\documentclass[a4paper,12pt]{article}
\usepackage{amsmath,amsfonts,amssymb,cite}

\begin{document}

\begin{center}
{\LARGE Generalized Soft Wall Model}
\end{center}

\begin{center}
{\large S. S. Afonin}
\end{center}

\begin{center}
{\small V. A. Fock Department of Theoretical Physics,
Saint-Petersburg
State University, 1 ul. Ulyanovskaya, 198504, Russia\\
Email: \texttt{afonin@hep.phys.spbu.ru} }
\end{center}

\begin{abstract}
We develop an exactly solvable generalization of the soft wall
holographic model for the vector mesons. The generalization
preserves the ultraviolet and infrared asymptotics of the soft wall
model and contains an additional free parameter. This new parameter
provides an arbitrary intercept in the Regge like spectrum of radial
excitations and leads to a substantial modification of asymptotic
expansion of the vector correlator at large momentum. The matching
to the Operator Product Expansion from QCD allows to estimate the
value of the new parameter which is shown to be in a good agreement
with the phenomenology. In addition, the mass splitting between the
vector and axial mesons arises naturally via the opposite sign of
the introduced contribution to the intercept.
\end{abstract}

\section{Introduction}

One of spectacular manifestations of confinement in QCD is the
expected Regge behavior of meson spectrum in the light quark sector,
both in the spin and in the radial directions. Traditionally the
related phenomenology was discussed in terms of the effective
strings or dual amplitudes. The story has taken an interesting turn
with the appearance of the Soft Wall (SW) holographic model for the
strongly coupled QCD~\cite{son2} which was inspired by the ideas of
the gauge/gravity correspondence in the string
theory~\cite{mald,witten,gub}. Since the SW model provided a seminal
theoretical setup for various applications, many authors proposed
different modifications of the SW model with the aim of improving it
or extending its applicability. We mention a few of them. The
relation of the SW model and the light-front QCD was studied in
Ref.~\cite{BT}. The revealed mapping of one approach onto the other
leads to interesting consequences concerning the hadron formfactors
and the dependence of hadron mass spectrum on the orbital quantum
number. The authors of Ref.~\cite{GKK} showed how to incorporate the
chiral symmetry breaking in the SW model in a way consistent with
QCD and with the known phenomenology. The two versions of the SW
model with a positive and negative dilaton profile for both mesons
and baryons of arbitrary spin were scrutinized in Ref.~\cite{GLSV}.
The SW model with the ultraviolet cutoff was analyzed in
Ref.~\cite{evans}. A incomplete list of the most recent developments
is given in Ref.~\cite{recent}.

In the original SW model, the intercept of linear in mass squared
trajectories of radially excited states is fixed. Naive attempts to
change it spoil the ultraviolet asymptotics of the model, as a
result the analytical behaviour of correlation functions becomes
strongly inconsistent with QCD. Namely, expanding the two-point
correlators at large Euclidean momentum generically there appear an
infinite number of terms which are absent in the standard Operator
Product Expansion (OPE). In the present paper, we show how to
introduce an arbitrary intercept in a self-consistent way. The
resulting model preserves the ultraviolet and infrared asymptotics
of the original SW model and the number of unwanted terms in the OPE
is reduced from infinity to one. We do not introduce any further
modifications. Such a SW model with arbitrary intercept is referred
to as generalized SW model. The arbitrary intercept entails some
important theoretical and phenomenological consequences which will
be analyzed.

The outline of this paper is as follows. The scheme of the vector SW
model is briefly reminded in Sect.~2. In Sect.~3, we derive our
generalization of this model. The two-point correlator and its
high-momentum expansion are calculated in Sect.~4. The Sect.~5
contains numerical fits and estimations. The Sect.~6 provides
further phenomenological discussions. We conclude in Sect.~7.

\section{Soft Wall Model}

In this section, we remind the reader the basic aspects of the SW
holographic model~\cite{son2}. For the sake of simplicity we will
consider the simplest Abelian version of this model that is defined
by the 5D action
\begin{equation}
\label{1}
S=\int d^4\!x\,dz\sqrt{g}\,e^{-az^2}\left(
-\frac{1}{4g_5^2}F_{MN}F^{MN}\right),
\end{equation}
where $g=|\text{det}g_{MN}|$, $F_{MN}=\partial_M V_N-\partial_N
V_M$, $M,N=0,1,2,3,4$, in the AdS background space whose metrics can
be parametrized as
\begin{equation}
\label{2}
g_{MN}dx^Mdx^N=\frac{R^2}{z^2}(\eta_{\mu\nu}dx^{\mu}dx^{\nu}-dz^2).
\end{equation}
Here $\eta_{\mu\nu}=\text{diag}(1,-1,-1,-1)$, $R$ denotes the AdS
radius, and $z>0$ is the holographic coordinate having the physical
sense of inverse energy scale. The boundary $z=0$ represents the 4D
Minkovski space.

The gauge invariance of the action~\eqref{1} permits to choose the
axial gauge, $V_z=0$, in which the equation of motion is simplified.
The spectrum $m_n$ of the physical vector mesons emerges from the
Kaluza-Klein decomposition of the field $V_{\mu}$,
\begin{equation}
\label{3}
V_{\mu}(x,z)=\sum_{n=0}^{\infty}V_{\mu}^{(n)}(x)v_n(z).
\end{equation}
For calculating the mass spectrum one should find the normalizable
solutions of the equation of motion for the 4D Fourier transform
$V_{\mu}^T(q,z)$ of the transverse components
($\partial^{\mu}V_{\mu}^T=0$). The normalizable eigenfunctions
$v_n(z)$ exist only for discrete values of 4D momentum
$q_n^2=m_n^2$. The corresponding equation that follows from the
action~\eqref{1} reads as follows
\begin{equation}
\label{4}
\partial_z\left(\frac{e^{-az^2}}{z}\partial_z v_n\right)+m_n^2\frac{e^{-az^2}}{z}v_n=0.
\end{equation}
This is a typical Sturm-Liuville problem. It is convenient to make
the substitution
\begin{equation}
\label{5}
v_n=\sqrt{z}e^{az^2/2}\psi_n,
\end{equation}
which transforms the Eq.~\eqref{4} in a Schr\"{o}dinger equation
\begin{equation}
\label{6}
-\psi_n''+V(z)\psi_n=m_n^2\psi_n,
\end{equation}
\begin{equation}
\label{7} V(z)=a^2z^2+\frac{3}{4z^2},
\end{equation}
where the prime stays for $\partial_z$. The eigenvalues of
Eq.~\eqref{6} yield the mass spectrum of the model
\begin{equation}
\label{8}
m_n^2=4|a|(n+1).
\end{equation}
Although the spectrum~\eqref{8} does not depend on the sign of $a$,
the choice $a<0$ leads to unphysical zero mode~\cite{son3}. For this
reason we will assume $a>0$ in what follows.

In a more general situation (for other spins) the
potential~\eqref{7} has two additional parameters
\begin{equation}
\label{9}
V(z)=a^2z^2+\frac{m^2-1/4}{z^2}+4ab,
\end{equation}
and results in the spectrum
\begin{equation}
\label{10}
m_n^2=2a(2n+m+1+2b).
\end{equation}

\section{Generalized Soft Wall Model}

We wish to derive an exactly solvable generalization of the vector
SW model that has an arbitrary intercept in the mass spectrum,
\begin{equation}
\label{11}
m_n^2=4a(n+1+b).
\end{equation}
Our generalization must not spoil neither ultraviolet (UV) nor infrared (IR)
asymptotics of the original SW model. We are going to show that this
requirement fixes unambiguously the form of the background in the 5D
action.

Let us write the holographic action in the form
\begin{equation}
\label{12}
S=\int d^4\!x\,dz f^2\left(-\frac{1}{4g_5^2}F_{MN}^2\right),
\end{equation}
with the unknown function $f(z)$ to be determined. The conformal
symmetry dictates the following UV asymptotics for this
function,
\begin{equation}
\label{13}
f(z)|_{z\rightarrow 0}\sim \frac{1}{\sqrt{z}}.
\end{equation}
If the condition~\eqref{13} is satisfied then in the UV limit the
action~\eqref{12} (written in the covariant form) has a form of the
action~\eqref{1}.

The equation for the mass spectrum is
\begin{equation}
\label{14}
\left(f^2v_n'\right)'+f^2m^2_nv_n=0.
\end{equation}
The substitution
\begin{equation}
\label{15}
v_n=\frac{\psi_n}{f},
\end{equation}
brings the Eq.~\eqref{14} into the form of a Schr\"{o}dinger
equation
\begin{equation}
\label{16}
-\psi_n''+\frac{f''}{f}\psi_n=m_n^2\psi_n.
\end{equation}
From~\eqref{10} it follows that for obtaining a shift in the
intercept the potential $\frac{f''}{f}$ must have the form
of~\eqref{9}. However, the choice $m^2\neq1$ will lead to a wrong UV
asymptotics in the vector SW model. The only possibility is to find
the function $f$ from the condition
\begin{equation}
\label{17}
\frac{f''}{f}=a^2z^2+\frac{3}{4z^2}+4ab,
\end{equation}
which has the form of Eq.~\eqref{6} with $m_n^2$ replaced by $m_n^2-4ab$.
This condition ensures the spectrum~\eqref{11} we are looking for.

The equation~\eqref{17} has two solutions --- an exponentially
decreasing and an exponentially growing one. To comply with the IR
asymptotics of the SW model (dictated by the absence of massless
mode and by the correct spectrum for the higher spin
mesons~\cite{son3}) we must select the decreasing solution.
Neglecting also the cases $b=-1,-2,\dots$ (since we do not want to
have any massless or tachyonic modes) the corresponding solution is
\begin{equation}
\label{18}
f=\Gamma(1+b)\frac{e^{-az^2/2}}{\sqrt{z}}U(b,0;az^2),
\end{equation}
where $U$ is the Tricomi confluent hypergeometric function and
$\Gamma$ is the Gamma function. We have chosen the normalization
$\frac{f^2}{z}=1$ at $z=0$.

Thus the action of the generalized SW model reads
\begin{equation}
\label{19}
S=\int d^4\!x\,dz\sqrt{g}\,e^{-az^2}U^2(b,0;az^2)\left(
-\frac{1}{4g_5^2}F_{MN}F^{MN}\right).
\end{equation}
This is our main result. Since
$U(b,0;0)=\Gamma^{-1}(1+b)=\text{const}$ and $U(b,0;az^2)\rightarrow
(az^2)^{-b}$ at $z\rightarrow\infty$, the obtained modification of
the 5D background does not affect neither UV nor leading IR
asymptotics. If for some reason one considers the SW model with
inverse dilaton background, $a<0$, the argument of the Tricomi
function must be changed to $|a|z^2$ (the function $U$ is complex at
negative argument).

It should be emphasized that the action~\eqref{19} is purely
phenomenological. The obtained background does not necessarily
follow from a dynamical solution of Einstein's equation. The model
can be also regarded as a compact five-dimensional writing of the
planar QCD sum rules in the spirit of the Ref.~\cite{rewr}.

\section{Vector Correlator}

The introduction of arbitrary shift in the spectrum brings
qualitatively new properties to the analytical structure of the
correlation functions. Following the standard recipe for the
holographic calculation of the correlators~\cite{witten,gub}, first
we should find the solution $v(q,z)$ ($q$ is the 4D momentum) of
equation of motion which is subject to the boundary condition
$v(q,0)=1$. For the action~\eqref{19} the corresponding solution is
\begin{equation}
\label{20}
v(q,z)=\frac{\Gamma\left(1+b-\tilde{q}^2\right)U\left(b-\tilde{q}^2,0;az^2\right)}
{\Gamma\left(1+b\right)U\left(b,0;az^2\right)},
\end{equation}
where the dimensionless momentum has been introduced, $\tilde{q}^2\equiv\frac{q^2}{4a}$.
The two-point correlation function of vector currents $J_{\mu}$,
\begin{equation}
\label{21}
\int d^4x e^{iqx}\langle J_{\mu}(x)J_{\nu}(0)\rangle=(q_{\mu}q_{\nu}-q^2g_{\mu\nu})\Pi_V(q^2),
\end{equation}
can be expressed via $v(q,z)$~\cite{son1,pom},
\begin{equation}
\label{22}
\Pi_V(q^2)=\left.\frac{R}{g_5^2}\frac{\partial_zv}{q^2z}\right|_{z\rightarrow 0}.
\end{equation}
The substitution of~\eqref{20} to~\eqref{22} gives the expression
\begin{equation}
\label{23}
\Pi_V(q^2)=\frac{R}{2g_5^2}\left\{-\psi\left(1+b-\tilde{q}^2\right)+
\frac{b}{\tilde{q}^2}\left[\psi\left(1+b-\tilde{q}^2\right)-\psi\left(1+b\right)\right]\right\},
\end{equation}
where $\psi$ denotes the digamma function. Applying the decomposition
\begin{equation}
\label{24}
\psi(1+x)=-\sum_{n=0}^{\infty}\frac{1}{x+n+1}+\text{const},
\end{equation}
we arrive at the spectral representation for the correlator under
consideration,
\begin{equation}
\label{25}
\Pi_V(q^2)=-\sum_{n=0}^{\infty}\frac{F_n^2}{q^2-4a(n+1+b)},
\end{equation}
\begin{equation}
\label{26}
F_n^2=\frac{2aR}{g_5^2}\left(1-\frac{b}{n+1+b}\right).
\end{equation}
The poles of the correlator yield the mass spectrum~\eqref{11}. At
$b\neq0$ the residues (they determine the electromagnetic decay
width, see~\eqref{30d}) acquire a dependence on $n$. Choosing $b<0$
this new feature allows to mimic a kind of the vector dominance in
the case of the $\rho$-mesons: The lightest vector state possesses
the largest value of residue. In the next Section, we demonstrate
that in the vector sector indeed $b<0$.

The expansion of the correlator~\eqref{23} at large Euclidean
momentum $Q^2=-q^2$ leads to
\begin{multline}
\left.\Pi_V\right|_{Q^2\rightarrow\infty}=\frac{R}{2g_5^2}\left\{\log\left(\frac{4a}{Q^2}\right)-
\frac{4a}{Q^2}\left[\frac12+b\left(\log\left(\frac{4a}{Q^2}\right)+1-\psi(1+b)\right)\right]\right.\\
+\left.\frac12\left(\frac{4a}{Q^2}\right)^2\left(\frac16-b^2\right)+
\frac16\left(\frac{4a}{Q^2}\right)^3b\left(b^2-\frac12\right)+\mathcal{O}\left(Q^{-8}\right) \right\}.
\label{27}
\end{multline}
The expansion~\eqref{27} can be matched to the Operator Product
Expansion for the vector two-point correlator~\cite{svz},
\begin{equation}
\label{28}
\Pi_V^{\text{(OPE)}}=\frac{N_c}{24\pi^2}\log\left(\frac{\mu_{\text{ren}}^2}{Q^2}\right)+
\frac{\alpha_s}{24\pi}\frac{\langle G^2\rangle}{Q^4}-
\frac{14\pi\alpha_s}{9}\frac{\langle\bar{q}q\rangle^2}{Q^6}+\mathcal{O}\left(Q^{-8}\right).
\end{equation}
The matching of coefficients  in front of the leading logarithm
provides the standard normalization factor,
\begin{equation}
\label{29}
\frac{R}{g_5^2}=\frac{N_c}{12\pi^2}.
\end{equation}

\section{Fits and Estimations}

In principle, the free parameters of the model --- the slope $4a$
and the (dimensionless) contribution to the intercept $b$ --- can be fixed by matching
the $\mathcal{O}(Q^{-4})$ and $\mathcal{O}(Q^{-6})$ terms. For the
typical phenomenological values of the gluon and quark condensates,
$\frac{\alpha_s}{\pi}\langle G^2\rangle=(360\,\text{MeV})^4$ and
$\langle \bar{q}q\rangle=-(235\,\text{MeV})^3$, one obtains
$4a=(905\,\text{MeV})^2 $ and $b=0.046$. Taking into account the
qualitative character of the model, these estimates look reasonable.

A more conservative point of view on the $\mathcal{O}(Q^{-6})$ term
would be to consider it as non-reliable for numerical fits because
of the asymptotic nature of the expansion. Within the standard SW
model, $b=0$, taking a typical phenomenological value for the slope
of meson trajectories~\cite{phen}, $4a\approx(1.1\,\text{GeV})^2$,
the matching of $\mathcal{O}(Q^{-4})$ terms in the
expansions~\eqref{27} and~\eqref{28} predicts a unrealistically
large value for the gluon condensate, $\frac{\alpha_s}{\pi}\langle
G^2\rangle\approx(440\,\text{MeV})^4$. The parameter $b$ allows to
remove this drawback: It can be fixed from the condition to have a
realistic gluon condensate in the expansions~\eqref{27},
\begin{equation}
\label{30}
b^2=\frac16-\frac{2\pi^2\frac{\alpha_s}{\pi}\langle G^2\rangle}{N_c(4a)^2}.
\end{equation}
Substituting the physical values for the slope and gluon condensate
to the condition~\eqref{30}, we arrive at the estimate
\begin{equation}
\label{30a}
|b|\approx 0.3.
\end{equation}
Below we show that this value is reasonable  from the
phenomenological point of view.

To compare the obtained estimate for the intercept parameter $b$
with the phenomenology we must make a fit of experimental masses by
the linear trajectory. The crucial point here consists in the choice
of data.
By construction, the model describes the isoscalar vector states,
{\it i.e.} we should consider the $\omega$-mesons in the vector
sector and the $f_1$-mesons in the axial-vector one. According to
the Particle Data~\cite{pdg}, there are only three well-established
$\omega$-mesons: $\omega(782)$, $\omega(1420)$, and $\omega(1650)$.
Taking their masses from~\cite{pdg} and ascribing them the "radial"
quantum numbers $n=0,1,2$, we obtain the fit (in GeV$^2$)
\begin{equation}
\label{30b}
m^2_{\omega}(n)\approx1.1(n+0.7).
\end{equation}
We do not write more accurate numbers because they would exceed the
accuracy of the large-$N_c$ limit --- typically about 10\%. In the
isoscalar axial-vector sector, there is only one well-established
state $f_1(1285)$ (another one, $f_1(1420)$, consists mostly of the
strange quarks). The best we can do is to use the non-confirmed
states $f_1(1970)$ and $f_1(2230)$~\cite{phen,pdg}. We ascribe them
the "radial" quantum numbers $n=2,3$ (the state corresponding to
$n=1$ --- the isoscalar partner of $a_1(1640)$~\cite{pdg} --- is not
known). This gives the fit
\begin{equation}
\label{30c}
m^2_{f_1}(n)\approx1.1(n+1.5),
\end{equation}
which should be regarded as a guess. It is quite remarkable that the
slopes in the linear spectra~\eqref{30b} and~\eqref{30c} are
approximately equal.

The spectrum~\eqref{30b} corresponds to $b\approx-0.3$
(see~\eqref{11}) that perfectly agrees with our estimate~\eqref{30a}
from the OPE. The data on axial-vector mesons seems to predict the
opposite sign for $b$. A very interesting feature of the generalized
SW model is that it allows the opposite signs of parameter $b$ for
the parity partners provided that the absolute value is the same. In
particular, the value $b=0.3$ may be compatible with the future data
because it leads to a spectrum which is (taking into account the
experimental errors) close to the guessed spectrum~\eqref{30c}.

A independent estimate for the parameter $b$ comes from the
calculation of electromagnetic decay width of vector mesons,
\begin{equation}
\label{30d}
\Gamma_{V\rightarrow e^+e^-}=C_V\frac{4\pi\alpha^2F_{V}^2}{3m_{V}},
\end{equation}
where $\alpha$ is the fine structure constant,
$\alpha=\frac{1}{137}$, and the quantity $F_{V}^2$ is given
by~\eqref{26} combined with~\eqref{29}. The factor $C_V$ reflects
the quark content of a given vector meson. Within the quark model,
the amplitude of electromagnetic decay is proportional to the
electric charge in the quark vertex. The quark content of the
$\rho^0$-meson is $\frac{u\bar{u}-d\bar{d}}{\sqrt{2}}$ and of the
$\omega$-meson is $\frac{u\bar{u}+d\bar{d}}{\sqrt{2}}$. As
$Q_u=\frac23$ and $Q_d=-\frac13$ one has $\Gamma_{\rho^0\rightarrow
e^+e^-}\sim(\frac23+\frac13)^2=1$ and $\Gamma_{\omega\rightarrow
e^+e^-}\sim(\frac23-\frac13)^2=\frac19$. By definition $C_{\rho}=1$,
hence $C_{\omega}=\frac19$. The value of $a$ in~\eqref{26} follows
from the fit~\eqref{30b}, $4a\approx1.1$~GeV$^2$.
Experimentally~\cite{pdg} $\Gamma_{\omega\rightarrow
e^+e^-}=0.60\pm0.02$~keV. The electromagnetic decay widths of the
excited $\omega$-mesons ($n>0$) and of $f_1$ are not known.
Substituting $n=0$ and $b=0$ we get $\Gamma_{\omega\rightarrow
e^+e^-}\approx0.4$~keV. The observable value is achieved if
$b\approx-0.3$ --- the same estimate as obtained above.

One can extend the analysis to the isovector case --- the $\rho$ and
$a_1$ mesons --- by considering the $SU(2)$ Yang-Mills field in the
action~\eqref{19}. If, as usual, the mass spectrum is defined by the
quadratic part of the holographic action, all formulas of the
previous Section will be the same. The ensuing numerical fits and
conclusions turn out to be very similar.

\section{Discussions}

The vector correlator of the SW model contains the
$\mathcal{O}(Q^{-2})$ term in the expansion at large Euclidean
momentum. Such a term is absent in the OPE~\eqref{28} by virtue of
the absence of dim2 local gauge-invariant operator in QCD (although
there are many speculations about the phenomenological relevance of
dim2 condensate~\cite{dim2}). Unfortunately, the generalized SW
model cannot solve this problem because of the logarithm in the
numerator of $\mathcal{O}(Q^{-2})$ term in the expansion~\eqref{27}.
More precisely, the problem can be partly resolved if one eliminates
the constant part in this numerator by fine-tuning the parameter $b$
(this would give $b\approx-0.24$). A possible physical origin of the
residual $\frac{\log{Q^2}}{Q^2}$ term in the OPE remains however
unclear. This problem seems to require a further modification of the
generalized SW model which is left for future.

The positivity of the $\mathcal{O}(Q^{-4})$ term in the OPE leads to
the constraint $|b|<1/\sqrt{6}$ in the expansion~\eqref{27}. In
addition, since this term is universal in the OPE for the vector and
axial-vector channels~\cite{svz} and depends quadratically on $b$,
one has an intriguing possibility for a self-consistent mass
splitting between the vector ($V$) and axial ($A$) states: The
corresponding spectra have universal absolute value of $b$ but
opposite sign,
\begin{equation}
\label{31}
m_{V,A}^2(n)=4a(n+1\mp|b|).
\end{equation}
As we have seen in the previous Section, this possibility seems to
be indeed realized in Nature with the absolute value $|b|\approx
0.3$.

A nonzero value of parameter $b$ generates the $\mathcal{O}(Q^{-6})$
term in the OPE and this represents a new feature in comparison with
the usual SW model. In the latter, the term $\mathcal{O}(Q^{-6})$ is
absent because the intercept (in units of the slope) is equal to
unity. It is well known~\cite{sr} that this is one of values of
intercept at which the term $\mathcal{O}(Q^{-6})$ disappears in the
OPE of the two-point correlators saturated by the narrow resonances
with linearly rising spectrum. It is interesting to note that in the
model~\eqref{31}, the $\mathcal{O}(Q^{-6})$ term in the $V$ and $A$
correlators differ by sign only (see Eq.~\eqref{27}). This is close
to the real OPE where these terms differ by the factor
$-\frac{7}{11}$~\cite{svz}. In this sense, the opposite sign of $b$
for the $V$ and $A$ mesons follows from the OPE. The factor
$-\frac{7}{11}$ can be reproduced only for different values of $b$
in the $V$ and $A$ sectors. But this would destroy the universality
of $\mathcal{O}(Q^{-4})$ term in the OPE which is related to the
gluon condensate. In view of the asymptotic nature of the OPE, the
$\mathcal{O}(Q^{-4})$ term is more reliable than the next
$\mathcal{O}(Q^{-6})$ one. For this reason we prefer to keep the
universality and use the $\mathcal{O}(Q^{-6})$ term for qualitative
conclusions at best. In our generalized SW model, the
$\mathcal{O}(Q^{-4})$ term in the OPE fixes the absolute value of
the intercept parameter $b$ and the $\mathcal{O}(Q^{-6})$ term
suggests the opposite sign for the $V$ and $A$ mesons. At the
present stage, we cannot derive the sign for say the $V$ mesons
theoretically. Only the phenomenology tells us that $b<0$ for the
$V$ mesons and $b>0$ for the $A$ states.

The SW model can be rewritten in an alternative form --- redefining
the vector field $V_M=e^{az^2/2}\tilde{V}_M$ in the action~\eqref{1}
leads to elimination of exponential background~\cite{nowall} (see
also~\cite{GLSV}). The price to pay is the appearance of an
effective potential, namely the $z$-dependent mass term,
\begin{equation}
\label{31b}
S=\int d^4\!x\,dz\sqrt{g}\left\{-\frac{1}{4g_5^2}\tilde{F}_{MN}\tilde{F}^{MN}+
\frac{a^2z^4}{2R^2g_5^2}\tilde{V}_M\tilde{V}^M\right\}.
\end{equation}
This mass term may be introduced in a gauge-invariant way via the
Higgs mechanism~\cite{nowall}: The action
\begin{equation}
\label{31c} S=\int
d^4\!x\,dz\sqrt{g}\left\{|D_M\varphi|^2-m_{\varphi}^2\varphi^2
-\frac{1}{4g_5^2}\tilde{F}_{MN}\tilde{F}^{MN}\right\},
\end{equation}
where $D_M=\partial_M-i\tilde{V}_M$ and the scalar field $\varphi$
is subjected to the free equation of motion in the AdS space,
\begin{equation}
\label{31d}
-\partial_z\left(\frac{\partial_z\varphi}{z^3}\right)+\frac{m_{\varphi}^2R^2\varphi}{z^5}=0,
\end{equation}
yields the action~\eqref{31b} if the Eq.~\eqref{31d} has the
solution $\varphi_0\sim z^2$, {\it i.e.} if the scalar mass is
$m_{\varphi}^2R^2=-4$. According to the AdS/CFT
dictionary~\cite{witten,gub}, the scalar mass is given by
$m_{\varphi}^2R^2=\Delta(\Delta-4)$, where $\Delta$ is the canonical
dimension of the corresponding scalar operator in CFT. In the case
under consideration, such a scalar field should be dual to a local
QCD operator of dimension two. In the light of this observation a
question appears how the generalized SW model looks like if we
redefine it in a similar way? Making the substitution
$V_M=e^{az^2/2}U^{-1}(b,0;az^2)\tilde{V}_M$ in the
action~\eqref{19}, we obtain
\begin{multline}
S=\int
d^4\!x\,dz\sqrt{g}\left\{-\frac{1}{4g_5^2}\tilde{F}_{MN}\tilde{F}^{MN}+\right.\\
\left.
\frac{a^2}{2R^2g_5^2}\left[z^2+2\frac{b}{a}-\frac{2U(b-1,0;az^2)}{aU(b,0;az^2)}\right]^2\tilde{V}_M\tilde{V}^M\right\}.
\label{31e}
\end{multline}
At $b=0$ the action~\eqref{31e} coincides with~\eqref{31b} since
$U(-1,0;az^2)=az^2U(0,0;az^2)$. If we now rewrite~\eqref{31e} in the
form of~\eqref{31c}, the solution for $\varphi$ leads to a
$z$-dependent mass $m^2_{\varphi}$. The physical interpretation
above emerges only in the deep infrared region,
$z\rightarrow\infty$. The leading contribution to $\varphi_0$ in
this region is $\varphi_0\sim z^2+b$ that would correspond to the 5D
mass $m_{\varphi}^2R^2=-\frac{4z^2}{z^2+b}$.

The extension of the SW model to the higher spin ($S$) mesons leads
to a nice relation $m_{n,S}^2=4a(n+S)$~\cite{son2} which is
compatible with other approaches (the Nambu-Goto strings, Veneziano
amplitudes). The background in the action~\eqref{19} does not lead
to a simple shift in the spectrum for higher $S$ (for the latter
purpose one would need a different background for each spin). One
can show that introducing the higher spin states according to the
scheme of Ref.~\cite{son2}, the potential~\eqref{17} of a
Schr\"{o}dinger equation in the background~\eqref{19} becomes
\begin{equation}
\label{32}
V(z)=a^2z^2+2a(S-1)+\frac{S^2-1/4}{z^2}+4ab\left[1+(S-1)\zeta(z)\right],
\end{equation}
where the function $\zeta(z)$,
\begin{equation}
\label{33}
\zeta(z)=\frac{U(1+b,1;az^2)}{U(b,0;az^2)},
\end{equation}
behaves as $\zeta(z\rightarrow0)\sim-\log{z}$ and
$\zeta(z\rightarrow\infty)\sim z^{-2}$ in the UV and IR limits.
Consequently the contribution due to nonzero $b$ does not affect the
UV and IR asymptotics of the potential~\eqref{32}. This contribution
will cause a slight deviation from linearity of the Regge like
spectrum.

\section{Conclusions}

We have shown how to introduce an arbitrary intercept to the linear
spectrum of the Soft Wall holographic model in a self-consistent
way. The spectrum of vector states becomes $m^2_n\sim n+1+b$, where
$n=0,1,2\dots$ and the case $b=0$ corresponds to the usual SW model.
The obtained generalization of the SW model remains exactly
solvable. The resulting freedom in the choice of the intercept
entails a sizeable modification of the two-point correlator,
specifically the residues of meson poles cease to be universal for
all states and a contribution related to the quark condensate is
generated. The latter signifies that the parameter $b$ is related to
the chiral symmetry breaking.

The Operator Product Expansion of two-point vector correlators
dictates the universal value of $|b|$ for the vector and
axial-vector particles but simultaneously indicates that the sign of
$b$ must be different in the vector and axial sectors. This
introduces the mass splitting between the vector and axial
particles. The phenomenological value of the gluon condensate and of
the slope of radial trajectories leads to the estimate
$|b|\approx0.3$. We studied the sector of isoscalar vector states.
The value $b\approx-0.3$ is in a perfect agreement both with the
well-established spectrum of $\omega$-mesons and with the
electromagnetic decay width of the $\omega(782)$-meson. The value
$b\approx0.3$ seems to be compatible with the spectrum of excited
axial $f_1$-mesons which is not yet well-established experimentally.

In view of recent phenomenological applications of the SW model and
attempts to derive it from a more fundamental setup, its
generalization presented in this work may be useful.

\section*{Acknowledgments}

I am grateful to Alexander Andrianov for many fruitful discussions
and to Domenec Espriu for the warm hospitality during my stay at the
University of Barcelona. The work is supported by the Dynasty
Foundation and by the RFBR grant 12-02-01121-a.

\end{document}